\documentclass[prl,superscriptaddress,twocolumn,showpacs,amsmath,%
preprintnumbers]{revtex4}
\usepackage{eucal} 
\usepackage{bm} 
\usepackage{graphicx}
\newcommand{\beq}{\begin{equation}}
\newcommand{\eeq}{\end{equation}}
\newcommand{\be}{\begin{eqnarray}}
\newcommand{\ee}{\end{eqnarray}}
\begin{document}
\title{Quantifying the nucleon's pion cloud with transverse charge densities}
\author{M.~Strikman}
\affiliation{Department of Physics, Pennsylvania State University,
University Park, PA 16802, USA}
\author{C.~Weiss}
\affiliation{Theory Center, Jefferson Lab, Newport News, VA 23606, USA}
\begin{abstract}
The transverse densities in a fast--moving nucleon offer a 
model--independent framework for analyzing the spatial structure 
of the pion cloud and its role in current matrix elements. 
We calculate the chiral 
large--distance component of the charge density using a dispersion
representation of the form factor and discuss its partonic interpretation.
The non--chiral core is dominant up to surprisingly large 
distances $\sim 2\, \text{fm}$. The chiral component 
can be probed in precision low--$Q^2$ elastic $eN$ scattering 
or in peripheral deep--inelastic processes which resolve 
its quark/gluon content.
\end{abstract}
%
%
%
%
\preprint{JLAB-THY-10-1169}
\maketitle
The large--distance behavior of strong interactions is governed 
by the spontaneous breaking of chiral symmetry in QCD, through 
which the pion appears as an almost massless Goldstone boson, 
coupling weakly to hadronic matter. The resulting effective
dynamics explains numerous observations in low--energy $\pi\pi$
and $\pi N$ scattering, the $NN$ interaction at large distances,
as well as weak and electromagnetic processes.
From the perspective of nucleon structure, these are often summarized 
in an intuitive spatial picture of the nucleon as consisting of a 
non--chiral ``core'' and a ``pion cloud'' 
of size $1/M_\pi$. Despite its widespread appeal and textbook--level 
status, this spatial picture has proved surprisingly 
difficult to quantify. Effective field theory (chiral perturbation 
theory, or ChPT) provides a systematic method to calculate 
dynamical effects at the scale $1/M_\pi$ but does not resolve the 
structure of the core, which is encoded in local counter 
terms \cite{Bernard:1995dp}.
Chiral soliton models of the nucleon provide a spatial picture of its
structure, but are restricted to the large--$N_c$ limit of QCD 
(semiclassical approximation) and subject to model assumptions 
about short--distance dynamics \cite{Brown:2010}. 
The lack of an unambiguous spatial representation
of the pion cloud is felt most acutely in elastic $eN$ scattering, 
which measures the charge and magnetization form factor of the 
nucleon. What is needed is a model--independent 
and fully quantitative formulation of the spatial structure of
the nucleon's chiral component appropriate for the analysis
of such measurements.

A new approach to this problem is possible with the recently 
proposed concept of transverse densities \cite{Miller:2007uy}.
Defined as 2--dimensional Fourier transforms of the elastic
form factors, they describe the distribution of charge and 
magnetization in the plane transverse to the direction of motion 
of a fast nucleon. In contrast to the traditional 
representation of form factors through 3--dimensional 
spatial densities in the Breit frame (zero energy 
transfer) \cite{Friedrich:2003iz,Hammer:2003qv}, 
the transverse densities provide an unambiguous spatial
interpretation also for systems in which the motion 
of the constituents is essentially relativistic. 
They are closely related to the parton picture of hadron 
structure in high--energy processes and correspond to 
a reduction of the generalized parton distributions (or GPDs)
describing the distribution of quarks/antiquarks with respect 
to longitudinal momentum and transverse position \cite{Belitsky:2005qn}.
In this way they establish an interesting connection between 
low--energy elastic $eN$ scattering and deep--inelastic processes 
sensitive to the transverse size of the nucleon, such as exclusive 
and diffractive processes in high--energy $eN$ and $NN$ 
scattering \cite{Frankfurt:2005mc}, and enable comprehensive
studies of the nucleon's spatial structure with several 
independent observables.

In this Letter we analyze the spatial structure of the nucleon's 
pion cloud and its role in elastic $eN$ scattering using the framework 
of transverse charge densities. We calculate the chiral 
component of the charge density in a $t$--channel representation 
of the form factor, which relates the large--distance behavior to 
the singularities in the timelike region.
It is shown that this formulation is equivalent to the partonic 
picture in the $s$--channel, where the large--distance behavior 
is governed by $\pi N$ and $\pi\Delta$ configurations 
in the nucleon's light--cone wave function. We find that the 
non-chiral core of the charge density is numerically dominant
up to surprisingly large distances $\sim 2 \, \text{fm}$,
and discuss the prospects for probing the chiral component
in precision low--$Q^2$ elastic scattering. A detailed account 
will be given in a forthcoming article.

The transverse charge density is defined as the 
2--dimensional Fourier transform of the Dirac form factor
of the vector current ($b \equiv |\bm{b}|$) \cite{Miller:2007uy}
\be
\rho (b) &=& \int \frac{d^2\Delta_\perp}{(2\pi)^2}
\; e^{-i (\bm{\Delta}_\perp \bm{b})}  \; F_1(t = -\bm{\Delta}_\perp^2) .
\label{rho_fourier}
\ee
In a frame where the nucleon is moving fast and the momentum 
transfer $\bm{\Delta}_\perp$ is in the transverse direction, 
$\bm{b}$ may be interpreted as the transverse position 
at which the current measures the charge density.
However, since the form factor is Lorentz--invariant, the 
quantity defined by Eq.~(\ref{rho_fourier}) may be evaluated
in more general ways.
In fact, the form factor $F_1(t)$ is an analytic function of the invariant
momentum transfer $t$, with singularities (branch cuts, poles) 
in the timelike region $t > 0$. By performing the angular integral
in Eq.~(\ref{rho_fourier}), and deforming the integration contour
in the radial variable $\Delta_\perp \equiv |\bm{\Delta}_\perp|$
to run along the imaginary axis,
one arrives at an equivalent representation of the transverse 
charge density as an integral over the imaginary part of the 
form factor in the timelike region ($K_0$ is the modified 
Bessel function):
\beq
\rho (b) \;\; = \;\; \int_{0}^\infty \frac{dt}{2\pi} 
\; K_0 (\sqrt{t} b)
\; \frac{\textrm{Im} F_1 (t + i0)}{\pi} .
\label{rho_timelike}
\eeq
It is particularly useful for discussing the asymptotic behavior of the 
charge distribution at large $b$. 
For a branch cut singularity at $t > \kappa^2$
with $\text{Im} F(t + i0) \propto (t - \kappa^2 )^\nu$ the charge density 
at large distances behaves as $\rho (b) \propto 
e^{-\kappa b}/(\kappa b)^{\nu + 3/2}$, with a coefficient which is
easily calculable from Eq.~(\ref{rho_timelike}). The exponent is 
determined by the position of the singularity only, while the 
pre-exponential factor depends on the power--like behavior
of the imaginary part near threshold. 
%
%
\begin{figure}
\includegraphics[width=.45\textwidth]{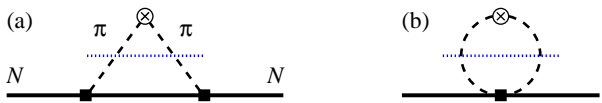}
\caption[]{Chiral processes contributing to the two--pion 
cut of the isovector nucleon form factor, or
the $\exp(-2M_\pi b)$ asymptotics of the transverse charge density.}
\label{fig:diag}
\end{figure}

\begin{figure}
\includegraphics[width=.48\textwidth]{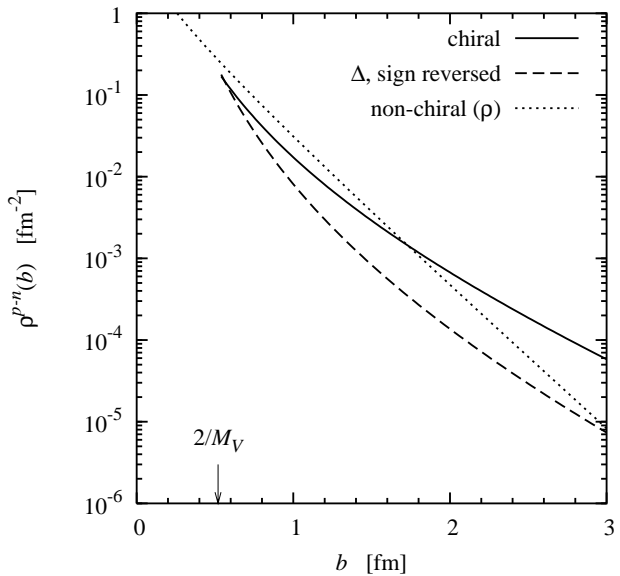}
\caption[]{Isovector transverse charge density in the nucleon, 
$\rho^{p - n}(b)$.
\textit{Solid line:} Chiral component from the processes 
of Fig.~\ref{fig:diag}, \textit{cf.}\ 
Eqs.~(\ref{rho_timelike}--\ref{imag_contact}).
\textit{Dashed line:} Contribution from $\Delta$ intermediate
states (sign reversed). {\it Dotted line:} Non-chiral 
charge density from $\rho$ exchange, \textit{cf.}\ Eq.~(\ref{rho_vmd}).}
\label{fig:vm}
\end{figure}

The leading singularity of the form factor in the timelike 
region is the two--pion cut at $t > 4 M_\pi^2$, which is 
of isovector nature. Because it lies in the unphysical region its
strength can only be calculated theoretically. In order to reliably
describe the charge density at $b \gtrsim 1/M_\pi$ we need an
approximation to the imaginary part which is generally 
accurate in the region $t - 4 M_\pi^2 \sim M_\pi^2$ 
and has the correct threshold
behavior in the limit $t \rightarrow 4 M_\pi^2$. 
It is provided by the amplitudes of Fig.~\ref{fig:diag},
where the $\pi$ and $N$ are pointlike and the couplings are 
those of the leading--order relativistic chiral Lagrangian 
of Ref.~\cite{Becher:1999he}. The loop integrals are evaluated
without expanding in the particle masses;
the resulting expression smoothly interpolates between 
the region $t - 4 M_\pi^2 \sim M_\pi^2$, where it contains
the leading term in heavy--baryon ChPT, and the near--threshold 
region where the heavy--baryon expansion does not converge; 
see Ref.~\cite{Becher:1999he} for a detailed discussion. 
Up to negligible terms of order
$t/M_N^2$, the result can be stated as (see also Ref.~\cite{Kaiser:2003qp})
\be
\frac{\textrm{Im} F_1^{p-n} (t + i0)}{\pi} 
&=& \frac{g_A^2 (t - 2 M_\pi^2)^2}{2 (4\pi F_\pi)^2 M_N \sqrt{t}}
(x - \arctan x) 
\phantom{xx}
\label{imag_pin} 
\\
&+& \frac{(1 - g_A^2) (t - 4 M_\pi^2)^{3/2}}{6 (4\pi F_\pi)^2 \sqrt{t}} ,
\label{imag_contact}
\ee
where $g_A = 1.26$ is the nucleon isovector axial coupling, 
$F_\pi = 93\, \text{MeV}$ the pion decay constant, and 
$x \equiv 2 M_N \sqrt{t - 4 M_\pi^2} / (t - 2 M_\pi^2)$.
The first term, Eq.~(\ref{imag_pin}), arises from the
diagram of Fig.~\ref{fig:diag}a and describes the contribution
of physical $\pi N$ intermediate state in the $s$--channel; 
the second term, Eq.~(\ref{imag_contact}), comes from the 
contact interaction of Fig.~\ref{fig:diag}b and the part
of diagram Fig.~\ref{fig:diag}a in which the nucleon pole 
is canceled. The transverse charge density resulting
from Eqs.~(\ref{rho_timelike}--\ref{imag_contact})
is shown in Fig.~\ref{fig:vm}. One sees that the $b$--dependence
is far from a simple exponential decay in the region shown
here, indicating strong variations in the pre-exponential factor.

%
%

It is interesting to study the large--distance behavior 
of the chiral component of the charge density obtained in this way.
Expanding Eqs.~(\ref{imag_pin}) and (\ref{imag_contact}) in powers 
of $\sqrt{t - 4 M_\pi^2}$, one gets an asymptotic series of the form
\beq
\rho^{p-n}(b) \;\; \propto \;\; \frac{e^{-2M_\pi b}}{(2 M_\pi b)^3} 
\left( 1 + \textrm{terms}\; \frac{M_N^2}{M_\pi^3 b} + \ldots \right) .
\eeq
The leading term in the pre-exponential factor dominates
only in the region $b \gg M_N^2/M_\pi^3$, corresponding to 
extremely large distances of the order $\sim 10^2 \, \text{fm}$;
already for $b \sim M_N^2/M_\pi^3$ the terms in the pre-exponential 
factor need to be summed up. The unusually slow convergence can be 
traced to an anomalous unphysical threshold close to $t = 4M_\pi^2$
in Eq.~(\ref{imag_pin}) \cite{Becher:1999he}; 
cf.\ also the discussion in Ref.~\cite{Hammer:2003qv}.
Alternatively, one may expand Eqs.~(\ref{imag_pin}) and (\ref{imag_contact})
in powers of $M_\pi/M_N$ in the region $t \sim M_\pi^2$, 
not necessarily close to threshold, corresponding to the leading term 
in heavy--baryon ChPT \cite{Becher:1999he}. This leads to an 
approximation for $\rho^{p-n}(b)$ valid in the region $b \sim 1/M_\pi$,
which, however, reproduces the finite--mass result only within a factor 
of $\sim 2$ in the region shown in Fig.~\ref{fig:vm}. Together,
these observations affirm the rationale of our interpolating approximation 
in numerical studies of the chiral component.

It is worth noting that in our dispersive approach the values 
of $t$ in Eq.~(\ref{rho_timelike}) are automatically
restricted to $\sqrt{t} \sim 1/b$, with exponential suppression
of large $t$. No external cutoff is needed. For values 
$b \sim 1/M_\pi$ the imaginary part is only sampled in a region 
where finite--size effects are negligible and it can safely be 
computed in the point particle approximation to chiral 
dynamics, \textit{cf.}\ the numerical studies in 
Ref.~\cite{Strikman:2009bd}. The transverse distance 
$b$ thus acts as an external parameter justifying the chiral expansion, 
similar to angular momentum in the partial--wave expansion of 
low--energy $\pi N$ scattering. 

Excitation of $\Delta$ resonances is known to play an important role 
in the two--pion cut; in particular, it ensures 
the proper scaling of the nucleon's 
isovector vector charge in the large--$N_c$ 
limit of QCD where $N$ and $\Delta$ become degenerate \cite{Cohen:1995mh}.
Using an empirical $\pi N\Delta$ coupling as in Ref.\cite{Strikman:2003gz},
it is straightforward to include the $\Delta$ in the process
of Fig.~\ref{fig:diag}a. We obtain an expression similar to
Eqs.~(\ref{imag_pin}) and (\ref{imag_contact}), which has opposite
sign and can be shown to cancel the $N$ diagram in the large--$N_c$ 
limit where $M_{N, \Delta} \sim N_c, M_\Delta - M_N \sim N_c^{-1},
t \sim M_\pi^2 \sim N_c^0$, and the 
couplings are simply related \cite{Strikman:2003gz}.
The transverse charge density from the $\Delta$ is shown in Fig.~\ref{fig:vm}. 
One sees that at $b < 1 \, \textrm{fm}$ it is comparable to that
from intermediate $N$ states, resulting in significant 
cancellations, but that at large $b$ the $N$ becomes numerically 
dominant.

To assess the numerical relevance of the chiral component at large $b$ 
we need to compare it to the bulk of the charge density unrelated 
to chiral dynamics. In the representation Eq.~(\ref{rho_timelike}), 
the latter is generated by the higher--mass singularities of the 
form factor at $t > 0$ and can be calculated in a systematic fashion. 
The leading one is the $\rho$ meson, which produces an 
asymptotic charge density
\beq
\rho^{p-n} (b)_{\text{$\rho$--exch.}} \;\; \sim \;\;
M_V^2 \, e^{-M_V b} / \sqrt{8\pi M_V b} ,
\label{rho_vmd}
\eeq
where $M_V = 770 \, \text{MeV}$ and the coupling has been chosen to 
ensure charge conservation  (we neglect the finite width).
Higher--mass states $\rho^{\prime}$ \textit{etc.} observed
in the photon spectral function, which build up the $1/t^4$ 
behavior of the form factor at large $|t|$,
are expected to have a negligible effect at the distance of interest here,
as is indeed confirmed by numerical studies.
Chiral and non--chiral components of the charge density are 
compared in Fig.~\ref{fig:vm}. One sees that the non-chiral component
is dominant up to distances $b \sim 2 \, \text{fm}$, the reason being 
the large coupling of the $\rho$ compared to two pions.
This result runs counter to naive expectations 
which place the region of the nucleon's ``pion cloud'' at
distances $> 1 \, \textrm{fm}$.

To what extent could present or future nucleon form factor measurements 
in the spacelike region $t < 0$ probe the chiral component in the 
transverse charge density? To answer this question, it is instructive
to consider the $b^2$--moments of the charge density, which are 
proportional to derivatives of the isovector form factor 
at $t = 0$:
\be
\langle b^{2n} \rangle &\equiv&
\int d^2 b \; b^{2n} \; \rho^{p-n} (b) 
\;\;\;\; (n = 1, 2)
\\
\langle b^{2} \rangle &=& 4F_1^{\prime}(0),
\;\;\;\; \langle b^{4} \rangle \;\; = \;\; 32 F_1^{\prime\prime}(0).
\ee
The moments of the chiral component can easily be computed by 
integrating Eq.~(\ref{rho_timelike}) over $b$. Restricting the
integral to values
$b > b_0 = \sqrt{\langle b^{2} \rangle}_{\text{$\rho$--exch.}} = 2/M_V$,
we find $\langle b^{2} \rangle_{\rm chiral}
= 0.08 \, \text{fm}^2$ (including both intermediate 
$N$ and $\Delta$), which amounts to only $14\%$ of the
experimental value $\langle b^{2} \rangle_{\text{exp}} = 
0.57 \, \text{fm}^2$. This value is consistent
with the uncorrelated two--pion contribution estimated in
the dispersion analysis of Ref.~\cite{Hammer:2003qv}.
Detecting the chiral component through an effect on the charge radius 
thus seems difficult. As an aside, we note that the expression for 
$\langle b^{2} \rangle$ obtained from 
Eqs.~(\ref{rho_timelike}--\ref{imag_contact}) formally reproduces the 
well--known divergence of the ``3--dimensional'' isovector charge 
radius in the chiral limit 
$M_\pi \rightarrow 0$ \cite{Cohen:1995mh,Beg:1973sc},
\beq
\langle r^{2} \rangle
\;\; \equiv \;\; \frac{3}{2} \langle b^{2} \rangle
\;\; \sim \;\; -\frac{(1 + 5 g_A^2) \ln M_\pi^2}{(4\pi F_\pi)^2} ;
\eeq
however, this is of little consequence for the numerical results
at the physical pion mass. A more promising chiral observable is
$\langle b^4 \rangle$, which receives most of its
contributions from distances $> 1\, \text{fm}$. 
We find $\langle b^4 \rangle_{\rm chiral} = 0.3 \, 
\text{fm}^4 \approx \langle b^2 \rangle^2_{\text{exp}}$
(including both $N$ and $\Delta$), a chiral contribution
comparable to the ``natural'' non-chiral value estimated from 
the charge radius. One should thus be able to see the chiral 
component in the behavior of the second derivative of the 
form factor at $t = 0$. The present form factor data at finite $t$ 
consistently extrapolate to $F_1(0) = 1$ with the slope (charge radius) 
measured in atomic physics experiments \cite{Eides:2007}, indicating 
that the second derivative could be extracted without tension 
given sufficiently precise data.

\begin{figure}
\includegraphics[width=.38\textwidth]{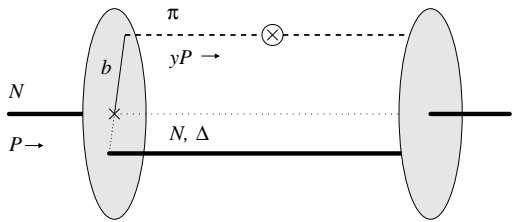}
\caption[]{Partonic interpretation of the chiral component of 
transverse charge density, cf.\ Fig.~\ref{fig:diag}. 
It corresponds to $\pi B \, (B = N, \Delta)$ configurations 
in the nucleon's light--cone wave function with pion momentum 
fraction $y \sim M_\pi / M_N$ and impact parameter $b \sim 1/M_\pi$.}
\label{fig:impact}
\end{figure}

The chiral large--distance component of the transverse charge density 
can also be discussed in the infinite--momentum frame, where it 
relates to the traditional concept of the ``pion cloud''
in the nucleon's partonic structure \cite{Sullivan:1971kd};
see Ref.~\cite{Strikman:2009bd} for a detailed discussion. 
Rewriting the invariant integrals for the form factor in terms 
of partonic variables, we find that the term Eq.~(\ref{imag_pin}) 
(coming from Fig.~\ref{fig:diag}a), and the corresponding one 
with the intermediate $\Delta$, can equivalently be expressed as
\beq
\rho^{p-n}(b) 
\;\; = \;\; \int_{0}^1 dy \, 
\left[ {\textstyle \frac{4}{3}} f_{\pi N}(y, b) 
- {\textstyle \frac{2}{3}} f_{\pi \Delta}(y, b) \right] ,
\label{rho_partonic}
\eeq
where $f_{\pi B}(y, b) \, (B = N, \Delta)$ are the 
distribution of pions in the fast--moving nucleon as a function 
of their longitudinal momentum fraction $y$ and transverse position 
$b$ \cite{Strikman:2009bd,Strikman:2003gz}. 
In this representation the chiral charge density arises as the 
cumulative effect of pions in the nucleon's light--cone wave function 
at transverse distances $b \sim 1/M_\pi$ --- a very intuitive 
picture, see Fig.~\ref{fig:impact}. 
The contact term in the form factor, Eq.~(\ref{imag_contact})
(coming from Figs.~\ref{fig:diag}a and b), when expressed 
in partonic variables, corresponds to a delta 
function--type contribution $f_{\pi B}(y, b)_{\text{contact}} \sim 
\delta (y) \, \rho(b)_{\text{contact}}$; such terms
indeed arise from the formal operator definition of 
the isovector pion momentum distribution in the nucleon.
At first sight, this appears to contradict the partonic 
interpretation of the transverse charge density in QCD, 
as only part of its chiral component seems to correspond to 
constituents carrying a finite fraction of the nucleon's momentum. 
The paradox is resolved when one realizes that 
the ``true'' pion density in the nucleon at small $y$ 
really involves high--mass intermediate states beyond the $N$ 
and $\Delta$. In a complete theory, the sum over such states would 
produce a smooth $y$--distribution of pions, whose integral then gives 
the chiral charge density at large $b$. In the present calculation 
based on point particles and the chiral Lagrangian, the effect 
of these high--mass states is summed up by contact terms.
In this sense, they represent legitimate contributions to the 
partonic charge density which are simply not ``resolved'' at the present 
level of approximations. (An explicit resummation of 
higher--mass multipion states was performed for a pion target within 
the leading logarithmic approximation of ChPT \cite{Kivel:2008ry}.)
Most of the nucleon's charge at large $b$ resides in the $\pi N$ 
component of the wave function, in which the contact term 
Eq.~(\ref{imag_contact}) 
contributes $\lesssim 10\%$ at $b > 1 \, \text{fm}$ 
(in the $\pi\Delta$ component it contributes about half). 
Thus, even at the present level of approximation, most of the
charge density at large $b$ can be directly interpreted in the 
partonic picture.
%
%

More generally, our 
approach suggests an interesting connection between low--energy 
elastic $eN$ scattering and the physics of peripheral high--momentum 
transfer processes in $eN, \gamma N$ and hadron--$N$ scattering,
which can resolve the partonic content of the pion cloud at a given
momentum fraction, $x$. Examples are exclusive meson/photon production
$eN \rightarrow e'N + M (M = \rho^0, \phi, J/\psi, \gamma)$ 
at $Q^2 \gg 1\, \text{GeV}^2$ and $|t| \ll 1\, \text{GeV}^2$, 
or corresponding processes in which the production happens on a 
pion at distances $b \sim 1/M_\pi$ which is knocked out and observed
in the final state \cite{Strikman:2003gz}. 
Such processes can measure the isoscalar 
quark and gluon density in the pion cloud, in which $N$ 
and $\Delta$ states contribute with the same sign and 
produce a sizable chiral component at 
$x < M_\pi / M_N$ \cite{Strikman:2003gz,Strikman:2009bd}.

In sum, the concept of transverse charge densities provides a rigorous
framework for analyzing the spatial structure of the nucleon's pion 
cloud and its contribution to current matrix elements. Our results 
quantify the impact of chiral dynamics on the analysis of empirical
charge densities \cite{Miller:2007uy,Rinehimer:2009sz}. 
The coordinate--space approach 
developed here can be extended to many other observables, 
such as magnetic form factors and the nucleon's orbital 
angular momentum content, gravitational form factors and 
the matter density \cite{Abidin:2008sb}, and 
$N\rightarrow \Delta$ transition form factors \cite{Carlson:2007xd}.

We are indebted to L.~Frankfurt, J.~Goity, N.~Kivel, G.~A.~Miller, 
and M.~V.~Polyakov for helpful discussions.
Notice: Authored by Jefferson Science Associates, LLC under U.S.\ DOE
Contract No.~DE-AC05-06OR23177. The U.S.\ Government retains a
non--exclusive, paid--up, irrevocable, world--wide license to publish or
reproduce this manuscript for U.S.\ Government purposes.
\end{document}